# CONTROLLING POSTURE USING A PLANTAR PRESSURE-BASED, TONGUE-PLACED TACTILE BIOFEEDBACK SYSTEM

Nicolas VUILLERME, Olivier CHENU, Jacques DEMONGEOT and Yohan PAYAN

Laboratoire TIMC-IMAG, UMR CNRS 5525, La Tronche, France

Address for correspondence:

Nicolas VUILLERME

Laboratoire TIMC-IMAG, UMR CNRS 5525

Faculté de Médecine

38706 La Tronche cédex

France.

Fax: (33) (0) 4 76 51 86 67

Email: nicolas.vuillerme@imag.fr

**Abstract**

The present paper introduces an original biofeedback system for improving human balance control, whose underlying principle consists in providing additional sensory information related to foot sole pressure distribution to the user through a tongue-placed tactile output device. To assess the effect of this biofeedback system on postural control during quiet standing, ten young healthy adults were asked to stand as immobile as possible with their eyes closed in two conditions of No-biofeedback and Biofeedback. Centre of foot pressure (CoP) displacements were recorded using a force platform. Results showed reduced CoP displacements in the Biofeedback relative to the No-biofeedback condition. The present findings evidenced the ability of the central nervous system to efficiently integrate an artificial plantar-based, tongue-placed tactile biofeedback for controlling control posture during quiet standing.

**Introduction**

Maintaining an upright stance posture involves the integration of sensory information from multiple sources including visual, somatosensory and vestibular systems (e.g. Massion 1994). One way to improve postural control is to give individual supplementary sensory information regarding their body's displacements and orientations, in addition to the above-mentioned usual sensory cues. Along these lines, numerous studies have reported that increased postural control could be obtained with the use of so-called biofeedback systems. Interestingly, most of these systems, widely used in physical therapy and rehabilitation, employed either the visual (e.g. Lajoie 2004, Shumway-Cook et al. 1988, Vaillant et al. 2004; Wu 1997) or auditory input (e.g. Chiari et al. 2005; Dozza et al. 2005; Hegeman et al. 2005, Petersen et al. 1996; Wong et al. 2001) to provide sensory information. At this point, however, these biofeedback systems, interfering ipso facto with the use of vision and hearing seem not particularly well-suited to applications in which users have to attend to several task simultaneously, but also for individuals with visual or hearing impairments. Within this context, the introduction of a tactile display, designed to evoke tactile sensation within the skin at the location of the tactile stimulator (e.g. Kaczmareck et al. 1991), using either mechanical ("vibrotactile display") (e.g. van Erp and van Veen 2006 ; Wall et al. 2001) or electrical ("electrotactile display") stimulation (e.g. Tyler et al. 2003), could present the advantage of freeing visual and auditory channels, by using another unexploited sensory channel to convey information about postural control.

Following this train of thought, we introduced an original biofeedback system for improving human upright balance control whose underlying principle consists in providing additional sensory information related to foot sole pressure distribution to the user through a tongue-placed tactile output device. The purpose of the present experiment was to assess the ability of the central nervous system to efficiently integrate this artificial biofeedback for controlling control posture during quiet standing.

**Methods**

<u>Subjects</u>

Ten young healthy adults (age: 24.6 ± 3.3 years; body weight: 71.5 ± 11.9 kg; height: 178.3 ± 10.0 cm; mean ± S.D.) were included in this study. They gave their informed consent to the experimental procedure as required by the Helsinki declaration (1964) and the local Ethics Committee, and were naive as to the purpose of the experiment. None of the subjects presented any history of motor problem, neurological disease or vestibular impairment.

<u>Task and procedures</u>

Subjects stood barefoot, foot together, their hands hanging at the sides, with their eyes closed. They were asked to sway as little as possible in two No-biofeedback and Biofeedback conditions. The No-biofeedback condition served as a control condition. In the Biofeedback condition, subjects performed the postural task using a plantar pressure-based, tongue-placed tactile biofeedback system. A plantar pressure data acquisition system (FSA Inshoe Foot pressure mapping system, Vista Medical Ltd.), consisting of a pair of insoles instrumented with an array of $8 \times 16$ pressure sensors per insole (1cm² per sensor, range of measurement: 0-600 mmHg), was used. The pressure sensors transduced the magnitude of pressure exerted on each left and right foot sole at each sensor location into the calculation of the positions of the resultant ground reaction force exerted on each left and right foot, referred to as the left and right foot centre of foot pressure, respectively ($CoP_{lf}$ and $CoP_{rf}$). The positions of the resultant CoP were then computed from the left and right foot CoP trajectories through the following relation (Winter et al. 1996):

$$CoP = CoP_{lf} \times R_{lf} / (R_{lf} + R_{rf}) + CoP_{rf} \times R_{rf} / (R_{rf} + R_{lf}),$$

where $R_{lf}$, $R_{rf}$, $CoP_{lf}$, $CoP_{rf}$ are the vertical reaction forces under the left and the right feet, the positions of the CoP of the left and the right feet, respectively.

CoP data were then fed back in real time to a recently developed tongue-placed tactile output device (Vuillerme et al. 2006a,b). This so-called Tongue Display Unit (TDU), initially introduced by Bach-y-Rita et al. (1998), comprises a 2D array (1.5 × 1.5 cm) of 36 electrotactile electrodes each with a 1.4 mm diameter, arranged in a 6 × 6 matrix. The matrix of electrodes, maintained in close and permanent contact with the front part of the tongue dorsum, was connected to an external electronic device triggering the electrical signals that stimulate the tactile receptors of the tongue via a flat cable passing out of the mouth. Note that the TDU was inserted in the oral cavity all over the duration of the experiment, ruling out the possibility the postural improvement observed in the Biofeedback relative to the No-biofeedback condition to be due to mechanical stabilization of the head in space. The underlying principle of our biofeedback system was to supply subjects with supplementary information about the position of the CoP relative to a predetermined adjustable "dead zone" (DZ) through the TDU. In the present experiment, antero-posterior and medio-lateral bounds of the DZ were set as the standard deviation of subject's CoP displacements recorded for 10 s preceding each experimental trial. A simple and intuitive coding scheme for the TDU, consisting in a "threshold-alarm" type of feedback rather that a continuous feedback about ongoing position of the CoP, was then used (Figure 1). (1) When the position of the CoP was determined to be within the DZ, no electrical stimulation was provided in any of the electrodes of the matrix (Figure 1, central panel). (2) When the position of the CoP was determined to be outside the DZ, electrical stimulation was provided in distinct zones of the matrix, depending on the position of the CoP relative to the DZ (Figure 1, peripheral panels). Specifically, eight different zones located in the front, rear, left, right, front-left, front-right, rear-left, rear-right of the matrix were defined ; the activated zone of the matrix corresponded to the position of the CoP relative to the DZ. For instance, in the case that the CoP was located towards the front of the DZ, a stimulation of the anterior zone of the matrix (i.e. stimulation of the front portion of the tongue) was provided (Figure 1, upper panel). Finally, in the present experiment, the frequency of the stimulation was maintained constant at 50 Hz across participants, ensuring a sensation of the continuous stimulation over the tongue surface. The intensity of the electrical stimulating current was adjusted for each subject, and for each of the front, rear, left, right, front-left, front-right, rear-left, rear-right portions of the tongue, given that the sensitivity to the electrotactile stimulation was reported to vary between individuals (Essick et al. 2003), but also as a function of location on the tongue in a preliminary experiment (Vuillerme et al. 2006a). Several practice runs were performed prior to the test to ensure that subjects had mastered the relationship between the position of the CoP relative to the DZ and lingual stimulations. Among all subjects, the observed maximum training time was below 5 min.

------------------------------------

Please insert Figure 1 about here

------------------------------------

A force platform (AMTI model OR6-5-1), which was not a component of the biofeedback system, was used to measure the displacements of the centre of foot pressure (CoP), as a gold-standard system for assessment of balance during quiet standing. Signals from the force platform were sampled at 100 Hz (12 bit A/D conversion) and filtered with a second-order Butterworth filter (10 Hz low-pass cut-off frequency).

Three 30s trials for each experimental condition were performed. The order of presentation of the two experimental conditions was randomized.

<u>Data analysis</u>

Two dependent variables were used to describe subject's postural behaviour. (1) The range of CoP displacements (mm) indicates the maximal deviation of the CoP in any direction. A large CoP range indicates that the resultant forces are displaced towards the balance stability boundaries of the participant and could challenge postural stability (e.g Patton et al. 2000). (2) The surface area (mm²) covered by the trajectory of the CoP with a 90% confidence interval (Tagaki et al. 1985) is a measure of the CoP spatial variability.

<u>Statistical analysis</u>

Data obtained for these two dependent variables were submitted to separate one-way analyses of variance (ANOVAs) (2 Conditions (No-biofeedback *vs.* Biofeedback)). Level of significance was set at 0.05.

**Results**

Figure 2 illustrates representative CoP displacements from a typical participant during standing in the No-biofeedback (2A) and Biofeedback (2B) conditions.

------------------------------------

Please insert Figure 2 about here

------------------------------------

Analysis of the range of the CoP showed a main effect of Condition, yielding a smaller range in the Biofeedback than No-Biofeedback condition ($F(1,9) = 20.60$, $P < 0.01$, Figure 2C). Similar results were obtained for the surface area covered by the trajectory of the CoP. The ANOVA confirmed the main effect of Condition, yielding a narrower surface area in the Biofeedback than No-Biofeedback condition ($F(1,9) = 15.72$, $P < 0.01$, Figure 2D).

**Discussion**

With the aim of improving balance control during quiet standing, the present experiment was designed to assess the postural effects of an original biofeedback system whose underlying principle consists in providing additional sensory information related to foot sole pressure distribution to the user through a tongue-placed tactile output device. To achieve this goal, young healthy adults were asked to stand as immobile as possible with their eyes closed in two conditions of No-biofeedback and Biofeedback. CoP displacements were recorded using a force platform.

Reduced range (Figure 2C) and surface area (Figure 2D) covered by the trajectory of the CoP were observed in the Biofeedback relative to the No-biofeedback condition. These results suggested the subjects were able to take advantage of an artificial tongue-placed tactile biofeedback to improve postural control during quiet standing. At this point, it is possible the sensory weighting of the tactile lingual cues for controlling posture to be subject of inter-individual variability. Indeed, such individual variations have previously been demonstrated for the degree to which subjects weight visual (e.g. Isableu et al. 1997), proprioceptive (e.g. Gurfinkel et al. 1995), somatosensory (e.g. Isableu and Vuillerme 2006) and vestibular (e.g. Horak and Hlavacka 2001) information for controlling their balance during quiet standing. An ongoing investigation involving a larger sample should allow us to address this issue in more depth. In general terms, however, our results suggest that an artificial tongue-placed tactile biofeedback can be efficiently integrated with other sensory cues by the postural control system to improve balance. These findings are in line with previous studies reporting that the availability of an augmented feedback - visual (e.g. Lajoie 2004; Shumway-Cook et al. 1988; Vaillant et al. 2004; Wu 1997) or auditory (e.g. Chiari et al. 2005; Dozza et al. 2005; Hegeman et al. 2005; Petersen et al. 1996; Wong et al. 2001) - yields postural improvement. However, as above-mentioned, a drawback of these biofeedback systems stem from the fact that they do interfere with vision or hearing. One possible solution to this problem is thus to distribute information across a dedicated sensory modality. For instance, using tactile, instead of visual or auditory output, would allow individuals perform several postural, visual and auditory tasks simultaneously. With this in mind, we developed an original biofeedback system comprising two major components: (1) the sensory unit and (2) the tongue-placed tactile output unit.

(1) A plantar pressure data acquisition system has been chosen as the sensory unit, since plantar cutaneous information is recognised to play a crucial role in the regulation of postural sway during quiet standing (e.g. Kavounoudias et al. 1998; Meyer et al. 2004). Indeed, plantar cutaneous mechanoreceptors could potentially provide detailed spatial and temporal information about contact pressures under the foot and shear forces resulting from body movement that constitute valuable feedback to the postural control system. In addition, it is important to mention that one of the more pervasive effects of aging is loss of cutaneous sensation (e.g. Kenshalo 1986; Skinner et al. 1984), which is known to correlate with impaired postural control and increased risk of falling (e.g. Lord et al. 1991; Tinetti and Speechley 1989). At this point, we believed that developing a biofeedback system designed to increase (and even substitute) somesthetic feedback provided by plantar soles would be beneficial to balance rehabilitation. Interestingly, interventions designated to enhance cutaneous sensation from the plantar soles resulting from therapeutic manipulation of the feet (Bernard-Demanze et al. 2004, 2006), noise-based techniques (Priplata et al. 2002, 2003) or the modification of the characteristics of the supporting surface on which individuals are standing previously have been shown to improve postural control (e.g. Maki et al. 1999; Okubo et al. 1980).

(2) The Tongue Display Unit (TDU), previously used for tactile-vision (e.g. Bach-y-Rita et al. 1969, 2003, Sampaio et al. 2001), tactile-proprioception (Vuillerme et al. 2006a,b) and tactile-vestibular sensory augmentation systems (Tyler et al. 2003), has been chosen as the sensory output unit. The performance of a tactile display largely depending on the neurophysiologic characteristics of the receptive body regions, the human tongue has recently been suggested to provide a promising electrotactile stimulation site (Bach-y-Rita et al. 1998). Indeed, because of its dense mechanoreceptive innervations (Trulsson and Essick 1997) and large somatosensory cortical representation (Picard and Olivier 1983), the tongue can convey higher-resolution information than the skin can (Sampaio et al. 2001; van Boven and Johnson 1994). In addition, due to the excellent conductivity offered by the saliva, electrotactile stimulation of the tongue requires only 3% of the voltage (5 - 15 V) and much less current (0.4 - 2.0 mA) than those required for the fingertip (Bach-y-Rita et al. 1998).

In summary, an original biofeedback system was developed for improving human balance control during quiet standing using plantar sensors and an electrotactile coupled via tongue human-machine interface. By showing reduced CoP displacements in the Biofeedback relative to the No-biofeedback condition, the present experiment evidenced the ability of the central nervous system to efficiently integrate an artificial plantar-based, tongue-placed tactile biofeedback for controlling control posture during quiet standing. From a fundamental perspective in the neurosciences area, these results, observed under "reliable" and "stable" sensory conditions in young healthy adults, encourage us to consider expanding future experiments to situations of sensory conflict, alteration, deprivation and/or restoration, to further investigate the re-weighting mechanisms involved in control of human posture. From a rehabilitative perspective, these results also could have implications for restoring balance control in individuals with reduced postural capacities, resulting either from normal aging, trauma or disease. Along these lines, the effectiveness of our biofeedback system in improving postural control in individuals with somatosensory loss in the feet from diabetic peripheral neuropathy and in persons with lower limb amputation is currently being evaluated.

**Acknowledgements**

The authors are indebted to Professor Paul Bach-y-Rita for introducing us to the TDU and for discussions about sensory substitution. The authors would like to thank subject volunteers. The company Vista Medical is acknowledged for supplying the FSA Inshoe Foot pressure mapping system. This research was supported by the Fondation Garches and the company IDS. Special thanks also are extended to P. Lashatte and F. Rerberthom for various contributions.

**References**

Bach-y-Rita P, Collins CC, Saunders F, White B, Scadden L (1969) Vision substitution by tactile image projection. Nature 221:963-964.

Bach-y-Rita P, Kaczmarek KA, Tyler ME, Garcia-Lara J (1998) Form perception with a 49-point electrotactile stimulus array on the tongue. J Rehabil Res Dev 35:427-430.

Bach-y-Rita P, Tyler ME, Kaczmarek KA (2003) Seeing with the brain. Int J Hum-Comput Int 15:285-295.

Bernard-Demanze L, Burdet C, Berger L, Rougier P (2004) Recalibration of somesthetic plantar information in the control of undisturbed upright stance maintenance. J Integr Neurosci 3:433-451.

Bernard-Demanze L, Vuillerme N, Berger L, Rougier P (2006) Magnitude and duration of the effects of plantar sole massages. International SportMed Journal. 7:154-169.

Chiari L, Dozza M, Cappello A, Horak FB, Macellari V, Giansanti D (2005) Audio-biofeedback for balance improvement: an accelerometry-based system. IEEE Trans Biomed Eng 52:2108-2111.

Dozza M, Chiari L, Horak FB (2005) Audio-biofeedback improves balance in patients with bilateral vestibular loss. Arch Phys Med Rehabil 86:1401-1403.

Essick GK, Chopra A, Guest S, McGlone F (2003) Lingual tactile acuity, taste perception, and the density and diameter of fungiform papillae in female subjects. Physiol Behav 80:289-302.

Gurfinkel VS, Ivanenko YP, Levik YS (1995) The influence of head rotation on human upright posture during balanced bilateral vibration. NeuroReport 7:137-140.

Hegeman J, Honneger F, Kupper M, Allum JH (2005) The balance control of bilateral peripheral vestibular loss subjects and its improvement with auditory prosthetic feedback. J Vest Res 15:109-117.

Horak FB, Hlavacka F (2001) Somatosensory loss increases vestibulospinal sensitivity. J Neurophysiol 86:575-585.

Isableu B, Ohlmann T, Crémieux J, Amblard B (1997) Selection of spatial frame of reference and postural control variability. Exp Brain Res 114:584-589.

Isableu B, Vuillerme N (2006) Differential integration of kinesthetic signals to postural control. Exp Brain Res DOI 10.1007/s0021-006-0630-4.

Kaczmareck KA, Webster JG, Bach-y-Rita P, Tompkins WJ (1991), Electrotactile and vibrotactile displays for sensory substitution systems. IEEE Trans Rehabil Eng 38:1-16.

Kavounoudias A, Roll R, Roll JP (1998) The plantar sole is a "dynamometric map" for human balance control. NeuroReport 9:3247-3252.

Kenshalo DR (1986) Somesthetic sensitivity in young and elderly humans. J Gerontol 41:632-642.

Lajoie Y (2004) Effect of computerized feedback postural training on posture and attentional demands in older adults. Aging Clin Exp Res 16:363-368.

Lord SR, Clark RD, Webster IW (1991) Physiological factors associated with falls in an elderly population. J Am Geriatr Soc 39:1194-1200.

Maki BE, Holliday PJ, Fernie GR (1990) Aging and postural control. A comparison of spontaneous- and induced-sway balance tests. J Am Geriatr Soc 38:1-9.

Maki BE, Perry SD, Norrie RG, McIlroy WE (1999) Effect of facilitation of sensation from plantar foot-surface boundaries on postural stabilization in young and older adults. J Gerontol A Biol Sci Med Sci 54:M281-M287.

Massion J (1994) Postural control system. Curr Opin Neurobiol 4:877-887.

Meyer PF, Oddsson LIE, De Luca CJ (2004). The role of plantar cutaneous sensation in unperturbed stance. Exp Brain Res 156:505-512.

Okubo J, Watanabe I, Baron JB (1980) Study on influences of the plantar mechanoreceptor on body sways. Agressologie 21:61-69.

Patton JL, Lee WA, Pai YC (2000) Relative stability improves with experience in a dynamic standing task. Exp Brain Res 135:117-126.

Petersen H, Magnusson M, Johansson R, Fransson PA (1996) Auditory feedback regulation of perturbed stance in stroke patients. Scand J Rehabil Med 28:217-223.

Picard C, Olivier A (1983) Sensory cortical tongue representation in man. J Neurosurg 59:781-789.

Priplata A, Niemi J, Salen M, Harry J, Lipsitz LA, Collins JJ (2002) Noise-enhanced human balance control. Phys Rev Lett 89:23101.

Priplata A, Niemi J, Salen M, Harry J, Lipsitz LA, Collins JJ (2003) Vibrating insoles and balance control in elderly people. Lancet 362:1123-1124.

Sampaio E, Maris S, Bach-y-Rita P (2001) Brain plasticity: 'visual' acuity of blind persons via the tongue. Brain Res 908:204-207.

Shumway-Cook A, Anson D, Haller S (1988) Postural sway biofeedback: its effect on reestablishing stance stability in hemiplegic patients. Arch Phys Med Rehabil 69:395-400.

Skinner HB, Barrack RL, Cook SD (1984) Age-related decline in proprioception. Clin Orthop 184:208-211.

Tagaki A, Fujimura E, Suehiro S (1985) A new method of statokinesigram area measurement. Application of a statistically calculated ellipse. In: Igarashi M, Black O (eds) Vestibular and Visual Control on Posture and Locomotor Equilibrium, Karger, Bâle, pp. 74-79.

Tinetti ME, Speechley M (1989) Prevention of falls among the elderly. N Engl J Med 320:1055-1059. Review.

Trulsson M, Essick GK (1997) Low-threshold mechanoreceptive afferents in the human lingual nerve. J Neurophysiol 77:737-748.

Tyler M, Danilov Y, Bach-y-Rita P (2003) Closing an open-loop control system: vetibular substitution through the tongue. J Integr Neurosci 2:159-164.

Vaillant J, Vuillerme N, Janvy A, Louis F, Juvin R, Nougier V (2004) Mirror versus stationary cross feedback in controlling the center of foot pressure displacement in quiet standing in elderly subjects. Arch Phys Med Rehabil 85:1962-1965.

van Boven RW, Johnson KO(1994) The limit of tactile spatial resolution in humans: grating orientation discrimination at the lips, tongue, and finger. Neurology 44:2361-2366.

Van Erp JBF, van Veen HAHC (2006) Touch down: the effect of artificial touch cues on orientation in microgravity. Neurosci Lett 404:78-82.

Vuillerme N, Chenu O, Demongeot J, Payan Y (2006a) Improving human ankle joint position sense using an artificial tongue-placed tactile biofeedback. Neurosci Lett 405:19-23.

Vuillerme N, Chenu O, Fleury, J, Demongeot J, Payan Y (2006b) Optimizing the use of an artificial tongue-placed tactile biofeedback for improving ankle joint position sense in humans 28th Annual International Conference of the IEEE Engineering In Medicine and Biology Society (EMBS), New York, USA.

Wall C, Weinberg MS, Schmidt PB, Krebs DE (2001) Balance prosthesis based on micromechanical sensors using vibrotactile feedback of tilt. IEEE Trans Biomed Eng 48:1153-1161.

Winter DA, Prince F, Frank JS, Powell C, Zabjek KF (1996) Unified theory regarding A/P and M/L balance in quiet stance. J Neurophysiol 75:2334-2343.

Wong MS, Mak AF, Luk KD, Evans JH, Brown B (2001) Effectiveness of audio-biofeedback in postural training for adolescent idiopathic scoliosis patients. Prosthet Orthot Int 25:60-70.

Wu G (1997) Real-time feedback of body center of gravity for postural training of elderly patients with peripheral neuropathy. IEEE Trans Rehabil Eng 5:399-402.

**Figure captions**

**Figure 1.** Sensory coding schemes for the Tongue Display Unit (TDU) as a function of the position of the centre of foot pressure (CoP) relative to a predetermined dead zone (DZ). Black triangles, dashed rectangles and black dots represent the positions of the CoP, the predetermined dead zones and activated electrodes, respectively. There were 9 possible stimulation patterns of the TDU. On the one hand, no electrodes were activated when the CoP position was determined to be within the DZ (central panel). On the other hand, 6 electrodes located in the front, rear, left, right, front-left, front-right, rear-left, rear-right zones of the matrix were activated when the CoP positions were determined to be outside the DZ, located towards the front, rear, left, right, front-left, front-right, rear-left, rear-right of the DZ, respectively (peripheral panels). These 8 stimulation patterns correspond to the stimulations of the front, rear, left, right, front-left, front-right, rear-left, rear-right portions of the tongue dorsum, respectively.

**Figure 2.** Representative displacements of the centre of foot pressure (CoP) from a typical subject recorded in the No-biofeedback (A) and Biofeedback (B) conditions. Mean and standard deviation of the range (C) and the surface area (D) of the centre of foot pressure (CoP) displacements obtained in the two No-biofeedback and Biofeedback conditions. These experimental conditions are presented with different symbols: No-biofeedback (*grey bars*) and Biofeedback (*black bars*). The significant *P*-values for comparison between No-biofeedback and Biofeedback conditions also are reported (\*\*: *P*<0.01).

**Figure 1.**

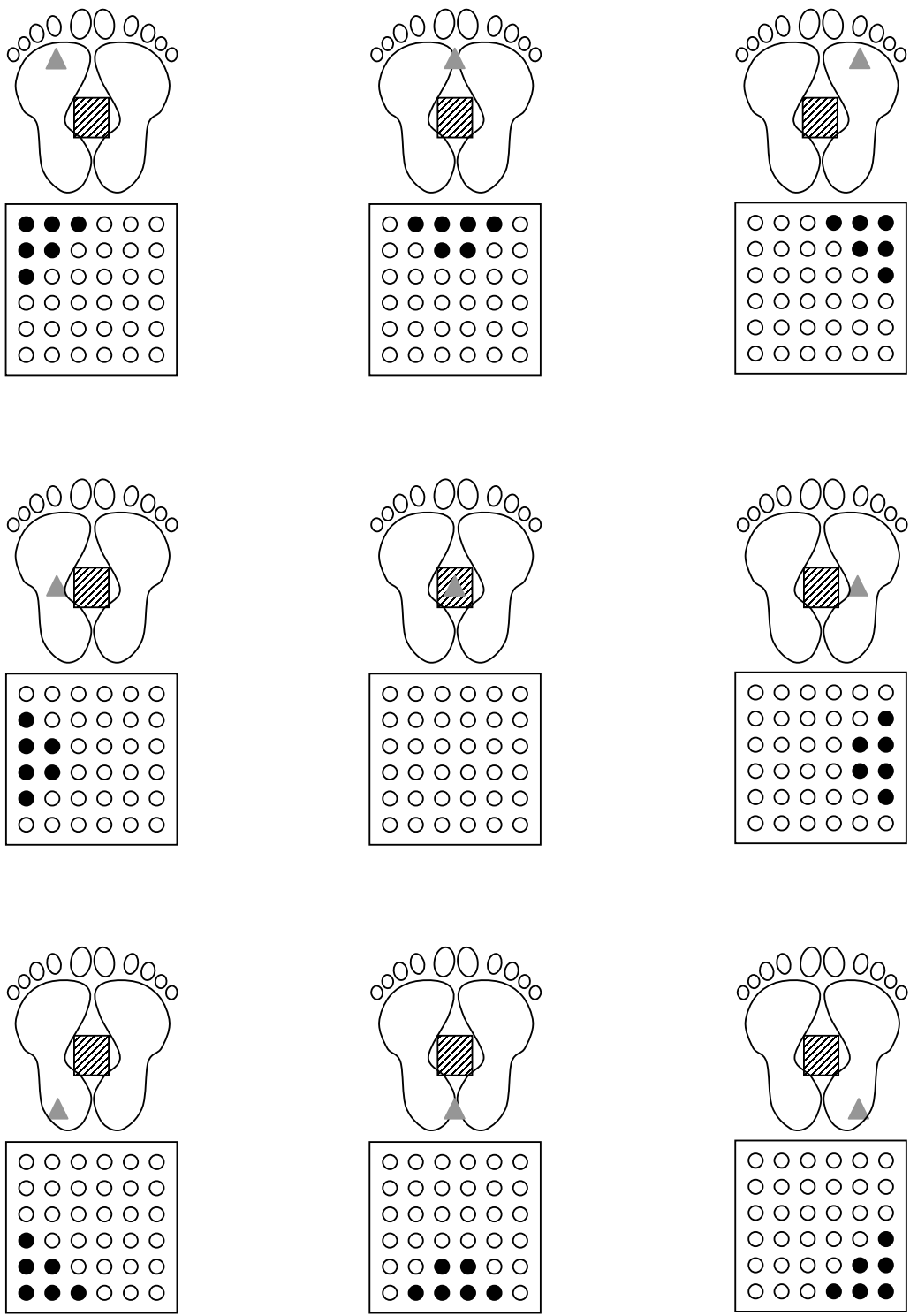

**Figure 2.**

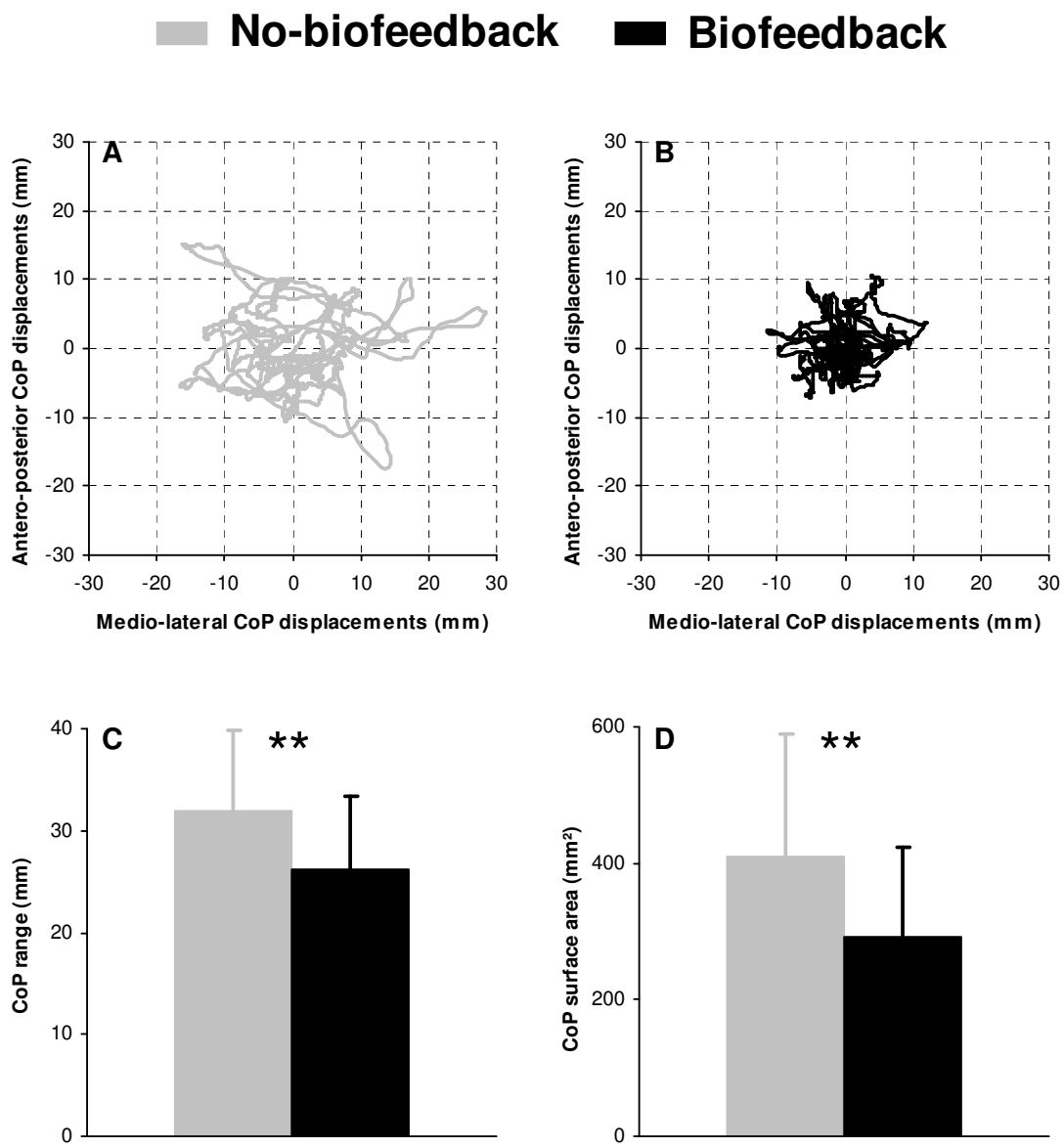